# Laser Offset Stabilization with Chip-Scale Atomic Diffractive Elements

Heleni Krelman[1], Ori Nefesh[1], Kfir Levi[1], Douglas G. Bopp[2,3], Songbai Kang[3], John E. Kitching[3] and Liron Stern[1]

[1]Institute of Applied Physics, Hebrew Univeristy of Jerusalem, Israel
[2]Vapor Cell Technologies, Boulder, CO, USA
[3]National Institute of Standards and Technology, Boulder, CO 80305, USA

Achieving precise and adjustable control over laser frequency is an essential requirement in numerous applications such as precision spectroscopy, quantum control, and sensing. In many such applications it is desired to stabilize a laser with a variable detuning from an atomic line. In this study, we employ an offset-stabilization scheme by utilizing phase contrast spectroscopy in microfabricated atomic diffractive elements vapor-cells. The spectroscopic response of such a device generates oscillating optical fringes, providing multiple optical frequency stabilization points across a bandwidth of tens of gigahertz, centered around the absorption resonances of Rb. Using this device, we demonstrate laser stabilization at various offset frequencies with instabilities reaching sub-megahertz levels. We further explore the fundamental limitations of our hybrid atomic-photonic device, drawing parallels to birefringent and dichroic spectroscopy apparatuses, which are commonly employed for offset stabilization. Our system showcases a broad offset lock bandwidth, a highly compact footprint, scalability to chip-scale production, and the ability to operate without reliance on magnetic fields. These attributes pave the way for a multitude of applications in quantum technologies.

## I. INTRODUCTION

Lasers are exquisite sources of monochromatic and intense light yet might drift and wander in frequency due to environmental perturbations. Control and stabilization of the frequency of a laser (often abbreviated 'locking') is typically achieved by using either optical frequency cavities[1], atomic or molecular systems[2–4], or a mixture of the two[5,6]. High stability and wide tunability of the laser frequency are required for several important applications in quantum technologies, such as quantum non-demolition experiments[7], Rydberg excitation[8], optical frequency metrology[9], and trapped atoms[10].

In pursuit of miniaturized and photonically integrated optical stabilization apparatuses, two different types of systems are being explored. First, optical cavities such as high-Q micro-ring resonator and compact Fabry-Pérot systems[6,11,12] are used to provide excellent short-term stability. In addition, they are relatively immune to frequency variations caused by power fluctuations, and allow tunable frequency locking points[13,14]. Yet, owing primarily to thermo-optic effects, cavity resonances drift in time. In contrast, atomic spectroscopic systems, which rely on the immutable and fundamental properties of atomic structure, can support highly stable and accurate optical signals[15,16]. Indeed, within the realm of chip-scale systems, micromachined vapor-cell-based optical frequency references and guided-wave photonically-integrated atomic systems have been reported demonstrating ultra-low levels of instabilities[6,17,18]. However, achieving tunability, i.e. offsetting the lock frequency, whilst maintaining stability of atomic lines in a compact platform is challenging.

Offset frequency locks are a class of atomic spectroscopy locks where the stabilized laser is detuned from the line center of an atomic resonance but derives its stability from the intrinsic atomic resonances through some form of transduction. Offset locking finds applications in various fields such as off-resonant optical spectroscopy[19], laser cooling[20], and imaging of ultracold atoms. Many manifestations of interference spectroscopy have been realized, including Dichroic Atomic Vapor Laser Lock (DAVLL), Faraday spectroscopy, and polarization-enhanced absorption spectroscopy (POLEAS). These methods involve the interference of two laser beams where one or both beams pass through an atomic gas and are differentially phase delayed and attenuated[17,21–25]. DAVLL and Faraday spectroscopy utilize this technique by passing two circularly polarized beams through an atomic vapor. The vapor exhibits optical activity in response to an applied magnetic field, shifting the absorption lines away from the exact atomic resonance via the Zeeman effect. The two beams are then interfered using a polarizing beam splitter and monitored with a photodetector. These spectroscopic signals can be manipulated by polarization control and differential attenuation; experimental realizations of this technique generate spectroscopic features over bandwidths that exceed 30 GHz using centimeter-scale, buffer-gas-filled vapor cells and Tesla-level magnetic fields. Such systems have shown megahertz-level stabilities over minute time-scales. However, the use of magnetic fields may be detrimental for some applications as it may influence nearby devices and experiments.

Recently, we have demonstrated micromachined atomic diffractive optical elements (ADOEs), which are in essence a system that geometrically confines atoms within micrometer-scale diffractive-optical elements[26]. In this manner, the atoms and the underlying material constitute a hybrid diffractive element, with a unique spectroscopic fingerprint which stems from this atomic-photonic coupling. In the context of laser-locking systems and in particular offset-lock systems, such hybrid photonic-atomic systems may support tunable atomic-derived offset locking by combining the above-mentioned advantages of stand-alone atomic and photonic systems.

Here, we present a hand-held hybrid atomic-photonic device that provides stable frequency offset locks. We demonstrate a simple, compact device that utilizes phase contrast spectroscopy of thermal Rb atoms in a micromachined, diffractive structure to generate a laser frequency reference with a bandwidth exceeding 70 GHz, and without requiring a magnetic field. We achieve frequency instabilities below $\approx 300$ kHz for integration times between 1 s and 500 s, reaching a minimal value of a few tens of kilohertz, limited by the residual amplitude modulation (RAM) of our current realization. We term this technique an Atomic Diffractive Laser Lock (ADLL).

## II. CONCEPT AND EXPERIMENTAL SYSTEM

We describe the basic concept of ADOEs, and how they are used for offset frequency locking. Fig. 1(a) illustrates a linear grating composed of an etched Si structure which is connected to an atomic reservoir. In essence, reflection of a beam from such a device can be viewed as a superposition of beams reflected from the surface and acquiring the appropriate phases. The reflection contains portions of the beam that acquire the phase of the atomic medium (referred to as "sample" in Fig. 1(a)), and portions that undergo regular reflection from Si surface ("reference"), resulting in the well-known interference structure in the far field. When combined with a high-density atomic dispersion (green curves in Fig. 1(a)) we achieve chirped sinusoidal fringes around the atomic absorption lines (see blue illustration in Fig. 1(a)). More generally, this spectroscopic arrangement yields signals that have a fingerprint originating from the interplay between the absorption and dispersion, i.e., the imaginary and real part of refractive index, respectively. Generally, when considering the optical frequency response of an atomic resonance for frequency detunings larger than the resonance linewidth, the dispersive nature of the resonances, as opposed to absorption, dominates the atomic response. This is a direct result of the atomic dispersion having a frequency dependence following $1/\omega$ in this regime, compared to atomic absorption, which has a $1/\omega^2$ dependence. The interference of two laser beams, where either one or both traverse an atomic medium, produces interference fringes which transduce the dispersion into chirped sinusoids. The periodicity of these sinusoids is directly linked to the frequency-dependent phase accumulation of light within the atomic vapor, particularly its frequency derivative. The latter effect directly correlates with the group index of the atomic vapor, also recognized as slow-light interferometric enhancement[27].

For a binary phase grating with a square cross section of a single groove, constituted from an etched frame with alternating columns of silicon and Rb atoms, the intensity at a given reflected non-zero diffraction order is proportional to $\sin^2(\Delta\phi/2)$. Here, $\Delta\phi = \Delta K \cdot 2L$ is the phase accumulation difference between portions of the beam interacting with rubidium atoms or reflecting from Si surface. $\Delta K$ is the wavenumber difference caused by interaction with the different refractive indices, while $L$ is the etch depth. We note that the portion of beam interacting with Rb traverses a distance $2L$ owing to our operation in the reflection mode. This equation demonstrates that the action of the diffraction grating is to convert the optical path length through the atomic vapor into an interferometer fringe as the laser frequency is scanned. Due to the common beam path, the only relevant length in this equation is the depth of the grating that holds the atomic vapor.

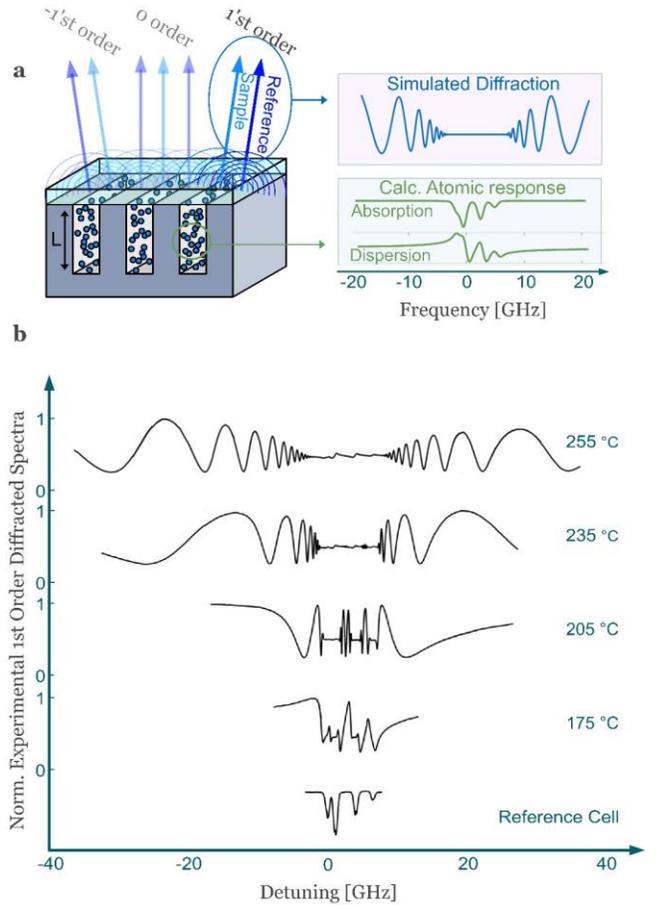

**FIG. 1.** (a) Schematic of the atomic grating illustrating the interference between beams that reflect off the top of the grating and that pass through the atomic medium prior to reflection. Illustrated angles are not to scale, as our first order diffraction angle is ≈0.6°. A calculated (blue line) atomic-grating response using the computed complex refractive index of Rb (green curves) is depicted to the right. (b) Experimental spectra of the first diffracted order across a temperature range of 175 °C to 255 °C, depicting the 70 GHz span of the high temperature spectra.

The phase difference contains the information about the refractive index of the atomic medium, which strongly depends on the temperature due to the variations in the atomic density. This results in a temperature dependent spectroscopic response, providing a broad spectrum of resonance frequencies. The position of a given fringe can be shifted by increasing the temperature of the whole atomic grating. Consequently, the bandwidth over which the fringes are observable is proportional to the optical density. This bandwidth can be altered by either varying the atomic density or the etch depth in the design. In Fig. 1(b) we present experimental spectroscopic results of our linear ADOE as a function of densities varying from ≈3×10$^{14}$ cm$^{-3}$ to ≈5×10$^{15}$ cm$^{-3}$, corresponding to temperatures of 175 °C to 255 °C respectively. At a temperature of 175 °C fringes primarily appear around and between the original absorption peaks. As the atomic density continues to rise, the spectrum showcases more oscillating patterns extending further from the absorption peaks. For instance, at a significantly elevated temperature of 255 °C, many fringes emerge, spanning a range exceeding 70 GHz. Yet, the high atomic density results in a flat-saturated optical response across the entire absorption manifold due to substantial atomic absorption. We note that the reflected light portion around this frequency position is directly related to Rb:Si duty cycle.

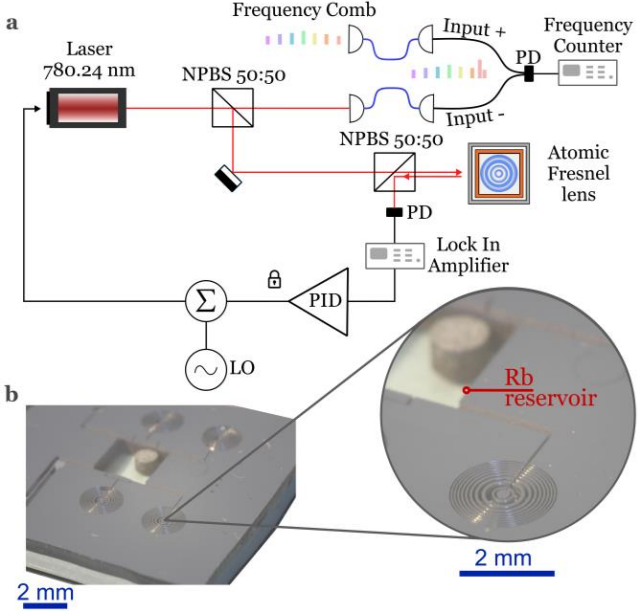

**FIG. 2.** (a) Illustration of the experimental arrangement for offset stabilization utilizing an ADOE. A modulated laser near resonance with the D2 line in Rb (780.24 nm) traverses through two 50:50 non-polarizing beamsplitters (NPBSs) and is reflected from a Fresnel lens at normal incidence; the diffracted focused beam is collected onto a photodetector (PD). To lock the laser, the collected light signal is demodulated to generate a dispersive error signal, which is fed to a PID controller. Sum of PID and local oscillator (LO) signals is injected into the laser. A portion of the light emitted from the stabilized laser creates a beat with a stabilized optical frequency comb to characterize the frequency stability. (b) Photograph of a microfabricated ADOE. The zoomed section showcases the Fresnel lens and connected Rb reservoir containing the dispenser pill.

Our chip-scale devices contain various types of microfabricated ADOEs (an example photograph is shown in Fig. 2(b)). In pursuing ADOE-based laser stabilization, we chose to work with an atomic Fresnel lens (AFL) whose spectral behavior is very similar to the above-mentioned linear-grating. Additionally, it supports high contrast and a simple collection of spectra as the interference signal appears at the single point, i.e., the focal distance of the lens. The lens (see zoomed photograph section in Fig. 2(b)), designed to have a focal length of 7 cm, has a diameter of 2 mm, etch depth of 150 μm, and distance between rings ranging from 40 μm to 120 μm. Operationally, we create the diffractive structures by using a silicon-on-insulator wafer, which is subsequently etched using deep reactive ion etching (DRIE). Here, we etch the silicon device layer down to the buried silicon dioxide layer, used as an etch stop. Next, a reservoir area is etched through the entire wafer thickness to connect to previously etched blind-hole channels. The micromachined Si wafer was anodically bonded to glass on both sides with a Rb-molybdate pill inside while under vacuum which results in a hermetically sealed AFL[18,28,29]. The devices were diced into 15 x 15 x 2 mm$^3$ chips containing four AFLs. The ADOEs presented here don't have any buffer gas and were fabricated in an evacuated environment of better than 10$^{-3}$ Pa (10$^{-5}$ mbar) background pressure.

The experimental setup is illustrated in Fig. 2(a). A laser beam of approximately 1 mm diameter and 20 μW of power irradiates the atomic lens at normal incidence with respect to the surface of the lens. To measure the reflected focused signal, a photodetector is placed following the NPBS at a distance close to the focal point. The wavelength of the laser (Rb D2 line of 780.24 nm) is modulated (i.e., its frequency is dithered [30]). To stabilize the laser, the signal from the detector is demodulated using a lock in amplifier, producing an error signal (ES). A Proportional-Integral-Derivative (PID) controller provides feedback to the laser based on the ES, thus stabilizing its frequency. A portion of the light emitted from the stabilized laser is heterodyned with an offset-stabilized optical frequency comb, whose repetition rate is referenced to a hydrogen maser, to characterize the frequency stability.

To ensure a stable temperature, we place our microfabricated cell in a custom-made double oven system. The outer layer, constructed from Al, is heated by resistor heaters, constituting the primarily heat source for the cell. The AFL is situated within a Cu layer, heated by a smaller heating resistor responsible for fine temperature adjustments. A Teflon layer in between facilitates a gradual dissipation of heat from the cell. Both heaters are regulated by PID controllers, enabling the achievement of millikelvin-level stability. During the long-term optical frequency measurement, we recorded the temperature of the AFL simultaneously. In the present study, we conducted experiments at a temperature of approximately 205 °C.

### III. RESULTS

The spectrum obtained at a temperature of 205 °C is shown in Fig. 3(a) (blue line). Frequency stabilization was carried utilizing four different peaks, denoted as I-IV. Peaks I, II, and III are situated between the reference absorption lines (Fig. 3(a) yellow line) and associated with a superposition of the refractive index tails of the surrounding absorption lines, i.e., the transitions between the $^{85}$Rb $5^2S_{1/2}$ (F=3) and $^{85}$Rb $5^2S_{1/2}$ (F=2) ground states to the $5^2P_{3/2}$ excited state manifold for peaks I and II, and transitions between the $^{85}$Rb $5^2S_{1/2}$ (F=2) and $^{87}$Rb $5^2S_{1/2}$ (F=1) ground states to the $5^2P_{3/2}$ excited states manifold, for peak III. The resonant linewidths of these peaks range from ≈200 MHz to 500 MHz. Peak IV stands apart from them, exhibiting a broader linewidth of around 1 GHz and associated with the slower decaying dispersive spectrum of the $^{87}$Rb $5^2S_{1/2}$ (F=1) to the $5^2P_{3/2}$ excited states manifold. By increasing the temperature by a few tens kelvin, it's possible to reduce the linewidth of all fringes, resulting in the generation of multiple lock points. For instance, the spectrum at 225 °C, shown in Fig. 3(a) as the green line, showcases lines with spectral widths ranging from ≈50 MHz to 500 MHz. The stabilized laser frequency was measured during a period of ≈ 1 hour for each stabilization point (Fig. 3(b)). All lock points exhibit megahertz-level stability while spanning a frequency range of several GHz.

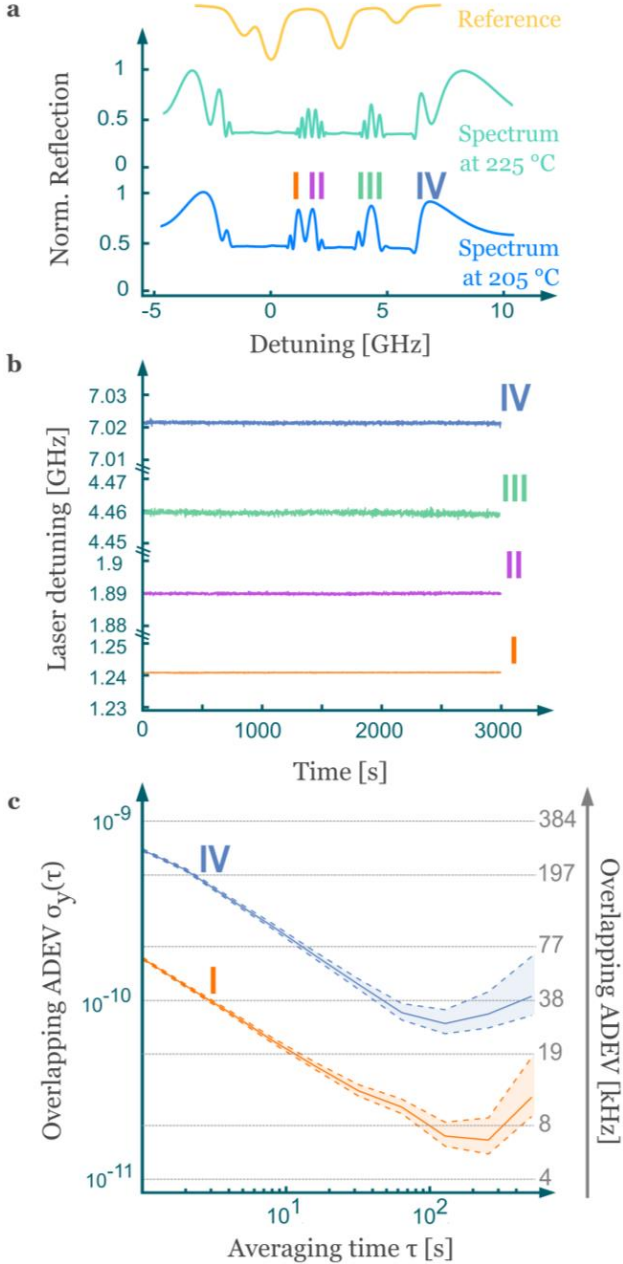

**FIG. 3.** (a) Spectral response of the AFL at 205 °C (blue line) and 225 °C. Labeling of peaks corresponds to those we use to perform optical-frequency stabilization. The yellow curve represents a reference spectrum of natural Rb. (b) Frequency as a function of time of a laser locked to the fringe peaks illustrated in Fig. 3(a). (c) The fractional frequency stability for peaks I and IV, illustrating stability below $8\times10^{-10}$ with a floor of ≈$1\times10^{-11}$, indicating sub-megahertz stability of the laser frequency at all measured times.

We further analyze the frequency stability statistical characteristics using overlapped Allan Deviation (ADEV). We focus on the two extremes of performance in our study, namely peaks I and IV (Fig. 3(c)). Both demonstrate a fractional frequency stability below $8\times10^{-10}$ at one second, i.e. below ≈300 kHz, with long-term stability below $10^{-10}$ out to 500 s, as low as $8\times10^{-11}$ (≈30 kHz) for the fourth peak and $2\times10^{-11}$ (≈10 kHz) for the first. This consistent behavior points to the laser's sub-megahertz stability at all measured times. For slightly longer integration times (≈3000 s) we observe an overall linear drift of $3.2\times10^{-10}$/h and $5.9\times10^{-10}$/h for the first and fourth peaks, respectively.

An initial and reasonable assumption posits that the ADEV of a narrower peak should demonstrate superior performance compared to that of a broader peak. While this holds true for peaks I and IV because of fivefold difference in linewidth, the ADEV for the fourth peak still achieves an unexpectedly good stability of $8\times10^{-10}$. We attribute this behavior to the unique shape of this specific peak, complicating the unequivocal determination of its width. Indeed, the frequency derivative of this line, which dictates the instability limit, has a sharper slope contributing to an enhanced ES.

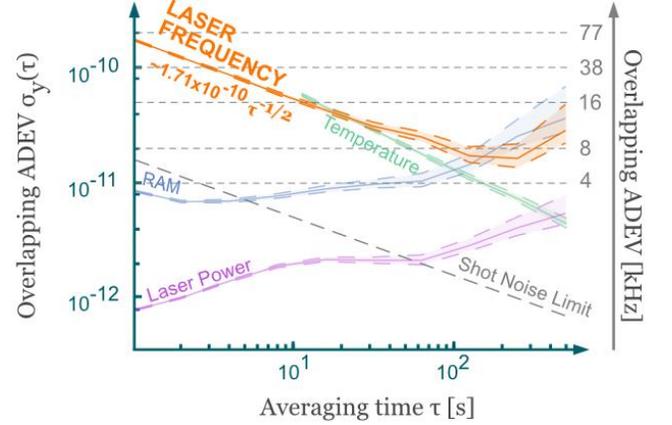

**FIG. 4.** The fractional frequency stability for peak I (orange line) and measured and calculated limitation factors: residual amplitude modulation (blue line), temperature (green line), laser power (purple line). Calculated shot noise is plotted by dashed gray line.

Our system is distinctive in leveraging the advantages of combining a photonic structure with an atomic medium. Consequently, there is an interest in examining the limitations inherent in such a hybrid system, particularly in comparison to typical standalone photonic and atomic subsystems. To that end, we study the limiting factors associated with our frequency stabilization platform by analyzing peak I. This peak exhibits the lowest fractional frequency instability, which at short time scales as $1.7\times10^{-10}\,\tau^{-1/2}$. Measurements of the amplitude noise around the modulation frequency are consistent with this short-term limit, and stem from technical intrinsic relative intensity noise (RIN) of our laser. Other limitations are also measured, such as RAM[31], temperature, and laser power, i.e., light shift. The corresponding frequency uncertainty for these factors are also plotted in Fig 4. We find that for averaging times longer than ≈10 s our measurements are limited by temperature fluctuations, whereas at ≈200 s (with a floor of nearly $2\times10^{-11}$) we are limited by RAM (blue line in Fig. 4) arising from the direct wavelength modulation of the laser.

Amongst the various aforementioned limitations, we carefully examine those that are inherently linked to the ADOE, namely the power coefficient and temperature coefficient (TC). These sensitivities are a characteristic of the absolute operating temperature, device geometry, and the nominal frequency detuning from the nearest resonances. The latter results from the fact that interference can be decomposed to a sum of differing refractive index "tails", each associated with an absorption line and a corresponding inherent sensitivity. To quantify these effects, we conducted direct measurements of thermal sensitivity by changing the temperature setpoint on the AFL while concurrently recording optical frequency measurements. The measured temperature coefficient (i.e., frequency deviation per unit temperature change) for this particular peak is approximately 10 MHz/K (averaged around the specific temperature working point), resulting in a $6.3\times10^{-11}$ fractional frequency instability per kelvin at 10 s, as illustrated in Fig 4 (green line). In this trace, relevant timescales larger than 10 s are plotted corresponding to the overall estimated thermal responsivity of our thermal management system. Other peaks also exhibit similar temperature dependence of order of few megahertz per kelvin. The experimental data agree with simulations and compare well to other forms of interference spectroscopy[32]. We have also performed similar measurements of the frequency dependence on incident power and have found a coefficient of ≈1.5 kHz/μW. It is intriguing to note that this coefficient is lower than the typical coefficient of a saturated absorption experiment, which is on the order of 10 kHz/μW[6,33]. We speculate that this lower light shift sensitivity, also observed previously[21], results from a relatively high atom-wall and atom-atom collision rate. These effects are attributed to the relatively high velocities of the relevant atomic velocity classes at such high frequency offsets.

Interestingly, whilst exploring the temperature dependence of peak I, we identified points where the coefficient changes its sign. We observed oscillations of the temperature-frequency coefficient, as well as a zero-crossing point at the working temperature. Physically, these oscillations stem from the superposition of different dispersions associated with the surrounding Rb transitions. Each line has a slightly different TC, such that the interference of multiple lines results in an oscillating overall TC. This suggests the presence of a working point where the frequency remains unaffected by temperature variations, which we have not exploited in this work.

To reduce the impact of RAM, which constitutes the primary limitation at longer time scales, we operated in a regime characterized by relatively low optical power and modulation amplitude. In the current configuration, we did not specifically control RAM or power. Introducing such control holds the potential to enhance frequency stability over longer periods by reducing the RAM and power contributions to the frequency instability. Simultaneously, it may be possible to work with higher modulation frequencies, potentially lowering the RIN. This approach, coupled with improved temperature stabilization, holds potential for improving both short- and long-term stabilities. In such a realization, convergence to the shot noise limit (depicted by the dashed gray line in Fig. 4) throughout all times could become achievable. We further note that this shot noise limit was calculated with respect to relatively low-power scenario we have described above, thus potentially providing room for improvement on the shot-noise limited metric. However, along with this improvement, the increase in power is expected to amplify the contribution of RAM and light shift to frequency instability, which may be mitigated by the above-mentioned active stabilization of RAM and power.

IV. DISCUSSION

The spectra from our interferometric system closely resemble those obtained using other offset-lock techniques such as DAVLL and in particular Faraday-based offset lock techniques. Indeed, Faraday rotation of a linearly-polarized laser beam, traversing through a circular birefringent Rb cell (i.e., subject to a longitudinal magnetic field) and subsequently transmitted through an analyzer (illustrated in Fig. 5), results in an intensity proportional to the difference or sum of the two decomposed fields in the circular-polarization basis. This is similar to the interference pattern obtained in our ADOEs, where the interference of reflected signals from the grating is the mechanism to produce the sum of two electric-field components as in Faraday spectroscopy. The advantage is that this occurs without the presence of circular birefringence, i.e., magnetic fields, resulting in an improved system in terms of size, weight and power (SWaP). The absence of magnetic fields is also advantageous, allowing the ADOE to be placed in close vicinity to magnetic-sensitive apparatuses, such as atomic systems.

The ADLL technique demonstrates large bandwidth and multiple reference frequency fringes, which can be easily tuned. On the one hand, tuning fringes that have ≈200 MHz width and separation between other fringes can be obtained by relatively small adjustments of the temperature (≈10 ºC). On the other hand, increasing the temperature even further can introduce new fringes and allows to extend the bandwidth to more than 70 GHz span.

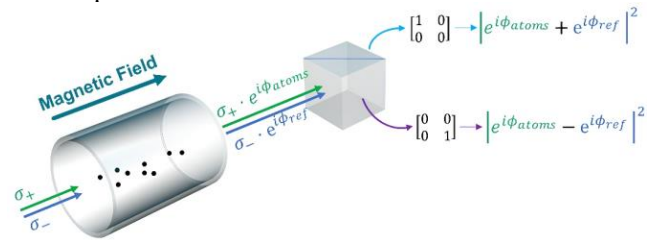

FIG. 5. Schematic illustration of Faraday spectroscopy. ADLL and Faraday techniques both operate under the same general principle of interference spectroscopy. In Faraday spectroscopy, right- and left-circular polarized components of the beam collect different phase due to the sensitivity of the atoms to light polarization, while ADLL achieves the same

effect by interfering portions of the incident beam which pass through the atomic medium or are reflected from Si.

Despite such strong coupling of our system to temperature, we observed temperature insensitive points, as is also seen in DAVLL systems[24]. Moreover, the location of these points may be tuned by modifying the linewidth of the atomic system, either by modifying the device geometry, changing the isotopic ratio, adding a buffer gas, or by all-optical alteration of the Rb index of refraction[26].

We can compare the different frequency sensitivities of our device to other common stabilization apparatuses. One important characteristic of frequency stabilization systems is the temperature coefficient. Low TC is necessary for minimizing temperature-induced drift and ensuring precision in optical measurements. The TC for the ADLL technique is $(\delta f/f)/T \approx 2.6 \times 10^{-8}$ 1/K, somewhat better than the TCs of a material-based Fabry-Pérot[34] or MRR cavity[35] ($\approx 10^{-6}$ 1/K and $\approx 2.5 \times 10^{-5}$ 1/K respectively) and a vacuum-based Fabry-Pérot cavity[36] ($\approx 5 \times 10^{-7}$ 1/K). More advanced systems involving ultra-low-expansion glass (ULE) cavities[37] exhibit a TC of about $10^{-9}$ 1/K. However, achieving sufficient thermal tunability requires a small free spectral range (FSR), which implies a large cavity size. Atomic systems involving saturated absorption spectroscopy[33] (SAS) in cm-scale cells have a TC $\approx 10^{-10}$ 1/K, but this stabilization technique doesn't allow wide bandwidth locking. Laser stabilization by sideband modulation does not depend on temperature, but the system itself is complex and requires high bandwidth modulators, synthesizers, and tight optical filters. Another essential feature of our system is the relatively low light-shift-induced frequency coefficient. The power coefficient of ADLL is measured to be 1.5 kHz/µW, while that of SAS is reported as $\approx 10$ kHz/µW and that of material-based cavities[6] is $\approx 200$ kHz/µW.

Additionally, an important figure of merit (FOM) for laser stabilization is the ratio between the peak's linewidth and contrast. The ADLL system demonstrates linewidths of $\approx 200$ MHz and typical contrast of $\approx 50\%$, so this FOM approaches that of techniques such as SAS and DAVLL. In general, the spectral response of the ADLL system can be modified by changing the geometry of the device, specifically the etch depth and width. A FOM for the tunable bandwidth and the stability can be enhanced in either parameter by changing the etch depth, etch width, temperature, or the absorption cross section. The latter can be modified by adding a buffer gas to broaden the optical transitions at the expense of reducing the total refractive index and absorption amplitude of the atoms. Finally, the contrast of the fringes is also related to the nonuniformity of the etch depth[26], which can be improved by optimizing the fabrication process.

## V. Conclusions

We have demonstrated experimentally a hybrid photonic-atomic platform allowing stable offset locks to the spectroscopic features of a chip-scale atomic diffractive optical element. The device utilizes concepts of interference spectroscopy, exploiting the slowly decaying dispersive wings of atomic vapors, to provide high-bandwidth optical fringes which can be used as discriminator signals for optical stabilization. By using a chip-scale atomic Fresnel lens, we demonstrate fractional optical frequency stabilities better than $8 \times 10^{-10}$ at short times and approaching $10^{-11}$ at longer times. The instability of these offset frequency locks is dominated by RIN at short times and by RAM at longer times. We discuss and measure the power and temperature dependence of such ADOEs and predict a shot-noise limited performance at the $10^{-11}$ level. The level of stability, miniaturization, scalability, simplicity, and absence of magnetic fields makes this system highly attractive. Moreover, the useable bandwidth containing optical fringes may exceed 70 GHz, with absolute frequency tunability that can be achieved using either relatively small temperature variations or photonic design. The concept of hybrid atomic-photonic laser stabilization merges the inherent advantages of both standalone atomic and photonic systems to provide enhanced capabilities, paving the way to enable a variety of quantum technologies.

**Commercial products disclaimer**

The full description of the procedures used in this paper requires the identification of certain commercial products. The inclusion of this information should in no way be construed as indicating that these products are endorsed by the National Institute of Standards and Technology (NIST) or are recommended by NIST or that they are necessarily the best materials for the purposes described.

**Acknowledgements**
The authors acknowledge Susan A. Schima for assistance in cell fabrication.